\documentclass[%
reprint,
showpacs,preprintnumbers,
nofootinbib,
amsmath,amssymb,
aps,
prd,
floatfix,
]{revtex4-1}

\usepackage{graphicx}
\usepackage{dcolumn}
\usepackage{bm}

\usepackage{url}

\newcommand{\boss}[2]{\ensuremath{\rlap{\kern-2.5pt\ensuremath{\overset{\scriptscriptstyle(-)}{\phantom{#1}}}}{\ensuremath{{#1}_{#2}}}}}

\begin{document}

\author{Carlo Giunti}
\email{giunti@to.infn.it}
\altaffiliation[also at ]{Department of Theoretical Physics, University of Torino, Italy}
\affiliation{INFN, Sezione di Torino, Via P. Giuria 1, I--10125 Torino, Italy}

\author{Marco Laveder}
\email{laveder@pd.infn.it}
\affiliation{Dipartimento di Fisica ``G. Galilei'', Universit\`a di Padova,
and
INFN, Sezione di Padova,
Via F. Marzolo 8, I--35131 Padova, Italy}

\date{\today}


\preprint{\begin{tabular}{l}
arXiv:1005.4599 [hep-ph]
\\
Phys.Rev.D82:053005,2010
\end{tabular}
}

\title{Short-Baseline Electron Neutrino Disappearance, Tritium Beta Decay and Neutrinoless Double-Beta Decay}

\begin{abstract}
We consider the interpretation of
the MiniBooNE low-energy anomaly
and
the Gallium radioactive source experiments anomaly
in terms of short-baseline electron neutrino disappearance
in the framework of 3+1 four-neutrino mixing schemes.
The separate fits of MiniBooNE and Gallium data
are highly compatible,
with close best-fit values of the effective oscillation parameters
$\Delta{m}^2$ and $\sin^2 2\vartheta$.
The combined fit gives
$\Delta{m}^2 \gtrsim 0.1 \, \text{eV}^2$
and
$
0.11
\leq \sin^2 2\vartheta \leq
0.48
$
at $2\sigma$.
We consider also the data of the Bugey and Chooz reactor antineutrino oscillation experiments
and the limits on the effective electron antineutrino mass in $\beta$-decay
obtained in the Mainz and Troitsk
Tritium experiments.
The fit of the data of these experiments
limits the value of $\sin^2 2\vartheta$ below
0.10
at $2\sigma$.
Considering the tension between
the neutrino MiniBooNE and Gallium data
and
the antineutrino reactor and Tritium data
as a statistical fluctuation,
we perform a combined fit
which gives
$\Delta{m}^2 \simeq 2 \, \text{eV}$
and
$
0.01
\leq \sin^2 2\vartheta \leq
0.13
$
at $2\sigma$.
Assuming a hierarchy of masses
$ m_{1}, m_{2}, m_{3} \ll m_{4} $,
the predicted contributions of $m_{4}$ to
the effective neutrino masses
in $\beta$-decay and neutrinoless double-$\beta$-decay are, respectively,
between about
0.06
and
0.49
and
between about
0.003
and
0.07
eV
at $2\sigma$.
We also consider the possibility of reconciling the tension between
the neutrino MiniBooNE and Gallium data
and
the antineutrino reactor and Tritium data
with different mixings in the neutrino and antineutrino sectors.
We find a
$2.6\sigma$
indication of a mixing angle asymmetry.
\end{abstract}

\maketitle

\section{\label{001}Introduction}

Neutrino oscillations have been observed in
solar, atmospheric and long-baseline reactor and accelerator experiments.
The data of these experiments are well fitted in the framework of
three-neutrino mixing,
in which the three flavor neutrinos
$\nu_{e}$,
$\nu_{\mu}$,
$\nu_{\tau}$
are unitary linear combinations of three massive neutrinos
$\nu_{1}$,
$\nu_{2}$,
$\nu_{3}$
with the solar (SOL) and atmospheric (ATM) squared-mass differences
\begin{align}
\null & \null
\Delta{m}^2_{21}
=
\Delta{m}^2_{\text{SOL}}
\simeq
8 \times 10^{-5} \, \text{eV}^2
\,,
\label{002}
\\
\null & \null
|\Delta{m}^2_{31}|
\simeq
|\Delta{m}^2_{32}|
=
\Delta{m}^2_{\text{ATM}}
\simeq
2 \times 10^{-3} \, \text{eV}^2
\,,
\label{003}
\end{align}
where
$\Delta{m}^2_{jk} = m_{j}^2 - m_{k}^2$
and
$m_{j}$ is the mass of the neutrino $\nu_{j}$
(see Refs.~\cite{hep-ph/9812360,hep-ph/0211462,hep-ph/0310238,hep-ph/0405172,hep-ph/0506083,hep-ph/0606054,GonzalezGarcia:2007ib,Giunti-Kim-2007}).

Besides these well-established observations of neutrino oscillations,
there are at least three anomalies which could be signals of
short-baseline neutrino oscillations
generated by a larger squared-mass difference:
the LSND
$\bar\nu_{\mu}\to\bar\nu_{e}$
signal \cite{hep-ex/0104049},
the Gallium radioactive source experiments anomaly \cite{0901.2200,1001.2731},
and the MiniBooNE low-energy anomaly \cite{0812.2243}.
In this paper we consider the MiniBooNE and Gallium anomalies,
which can be explained by short-baseline electron neutrino disappearance
\cite{0707.4593,0711.4222,0902.1992}
in the effective framework of four-neutrino mixing,
as explained in Sections~\ref{009} and \ref{019}.
On the other hand,
the LSND anomaly is disfavored by the results of the MiniBooNE
$\nu_{\mu}\to\nu_{e}$
experiment
\cite{0704.1500,0812.2243}
and may require another explanation
\cite{hep-ph/0010308,0705.0107,0706.1462,0710.2985,0805.2098,0906.1997,0906.5072}.

In Refs.~\cite{0707.4593,0902.1992}
we proposed to explain the MiniBooNE
low-energy anomaly \cite{0704.1500,0812.2243}
through the disappearance of electron neutrinos
due to very-short-baseline oscillations into sterile neutrinos
generated by a squared-mass difference $ \Delta{m}^2 $ larger than about $20\,\text{eV}^2$.
In that case,
the analysis of the MiniBooNE data is simplified by the fact that
the effective survival probability $P_{\nu_{e}\to\nu_{e}}$ is practically constant in the MiniBooNE energy range
from 200 to 3000 MeV.
In this paper we extend the analysis of MiniBooNE data to lower values of $ \Delta{m}^2 $,
considering the resulting energy dependence of the effective
short-baseline (SBL) electron neutrino and antineutrino survival probability
\begin{equation}
P_{\boss{\nu}{e}\to\boss{\nu}{e}}^{\text{SBL}}(L,E)
=
1
-
\sin^2 2\vartheta
\sin^2\!\left( \frac{ \Delta{m}^2 L }{ 4 E } \right)
\,,
\label{004}
\end{equation}
where $L$ is the neutrino path length and $E$ is the neutrino energy
(CPT invariance implies that the survival probabilities of
neutrinos and antineutrinos are equal;
see Ref.~\cite{Giunti-Kim-2007}).

The two-neutrino-like effective short-baseline survival probability in Eq.~(\ref{004}) is obtained
in four-neutrino schemes
(see Refs.~\cite{hep-ph/9812360,hep-ph/0405172,hep-ph/0606054,GonzalezGarcia:2007ib}),
which are the simplest extension of three-neutrino mixing schemes which can accommodate
the two small solar and atmospheric squared-mass differences in Eqs.~(\ref{002}) and (\ref{003}),
and one larger squared-mass difference for short-baseline neutrino oscillations,
\begin{equation}
|\Delta{m}^2_{41}| = \Delta{m}^2 \gtrsim 0.1 \, \text{eV}^2
\,.
\label{005}
\end{equation}
The existence of a fourth massive neutrino corresponds,
in the flavor basis,
to the existence of a sterile neutrino $\nu_{s}$.

In this paper we consider 3+1 four-neutrino schemes,
since 2+2 four-neutrino schemes are disfavored by the combined constraints
on active-sterile transitions in solar and atmospheric neutrino experiments
\cite{hep-ph/0405172}.
For simplicity,
we consider only 3+1 four-neutrino schemes with
\begin{equation}
m_{1}, m_{2}, m_{3} \ll m_{4}
\,,
\label{006}
\end{equation}
which
give the $\Delta{m}^2_{41}$ in Eq.~(\ref{005})
and
appear to be more natural than the other possible
3+1 four-neutrino schemes in which
either three neutrinos or all four neutrinos
are almost degenerate at a mass scale larger than
$\sqrt{\Delta{m}^2}$
(see Refs.~\cite{hep-ph/9812360,hep-ph/0405172,hep-ph/0606054,GonzalezGarcia:2007ib}).

In 3+1 four-neutrino schemes the effective mixing angle in the
effective short-baseline electron neutrino survival probability in Eq.~(\ref{004})
is given by
(see Refs.~\cite{hep-ph/9812360,hep-ph/0405172,hep-ph/0606054,GonzalezGarcia:2007ib})
\begin{equation}
\sin^2 2\vartheta
=
4 |U_{e4}|^2 \left( 1 - |U_{e4}|^2 \right)
\,.
\label{007}
\end{equation}
In this paper we assume
that the value of $|U_{\mu4}|^2$ is so small that the
effective short-baseline muon neutrino survival probability is practically equal to unity
and
short-baseline $\boss{\nu}{\mu}\leftrightarrows\boss{\nu}{e}$
are negligible\footnote{
In 3+1 four-neutrino schemes the effective
short-baseline muon neutrino survival probability
has the form in Eq.~(\ref{004}) with
$\sin^2 2\vartheta$
replaced by
$\sin^2 2\vartheta_{\mu\mu} = 4 |U_{\mu4}|^2 \left( 1 - |U_{\mu4}|^2 \right)$.
The effective
short-baseline $\boss{\nu}{\mu}\leftrightarrows\boss{\nu}{e}$
transition probability is given by
$
P_{\boss{\nu}{\mu}\leftrightarrows\boss{\nu}{e}}^{\text{SBL}}(L,E)
=
\sin^2 2\vartheta_{e\mu}
\sin^2\!\left( \frac{ \Delta{m}^2 L }{ 4 E } \right)
$,
with
$\sin^2 2\vartheta_{e\mu} = 4 |U_{e4}|^2 |U_{\mu4}|^2$
(see Refs.~\cite{hep-ph/9812360,hep-ph/0405172,hep-ph/0606054,GonzalezGarcia:2007ib}).
}.
This assumption is justified by the lack of any indication of
$\nu_{\mu}\to\nu_{e}$
transitions in the MiniBooNE experiment
\cite{0704.1500,0812.2243}
and the limits on short-baseline muon neutrino disappearance
found in the
CDHSW \cite{Dydak:1984zq},
CCFR \cite{Stockdale:1985ce} and
MiniBooNE \cite{0903.2465}
experiments.
We do not consider the MiniBooNE antineutrino data
\cite{0904.1958},
which have at present statistical uncertainties which are too large to constraint new physics
\cite{0902.1992}.

The plan of the paper is as follows.
In Section~\ref{009} we discuss the analysis of MiniBooNE data.
In Section~\ref{019} we present an update of the analysis of Gallium data
published in Ref.~\cite{0711.4222}
and the combined analysis of
MiniBooNE and Gallium data.
In Section~\ref{027} we discuss the implications of the
measurements of the effective electron neutrino mass in Tritium $\beta$-decay experiments
and their combination with reactor neutrino oscillation data.
In Section~\ref{041}
we present the results of the combined analysis of MiniBooNE, Gallium, reactor and Tritium data
and
in Section~\ref{045} we present the corresponding predictions for the effective masses measured in
$\beta$-decay and neutrinoless double-$\beta$-decay experiments.
In Section~\ref{058} we calculate the mixing angle asymmetry
between the neutrino and antineutrino sectors which could explain the
tension between the neutrino and antineutrino data
under our short-baseline $\nu_{e}$-disappearance hypothesis.
In Section~\ref{060} we draw the conclusions.

\begin{figure*}
\begin{center}
\begin{tabular}{lr}
\includegraphics*[bb=33 147 578 695, width=0.48\linewidth]{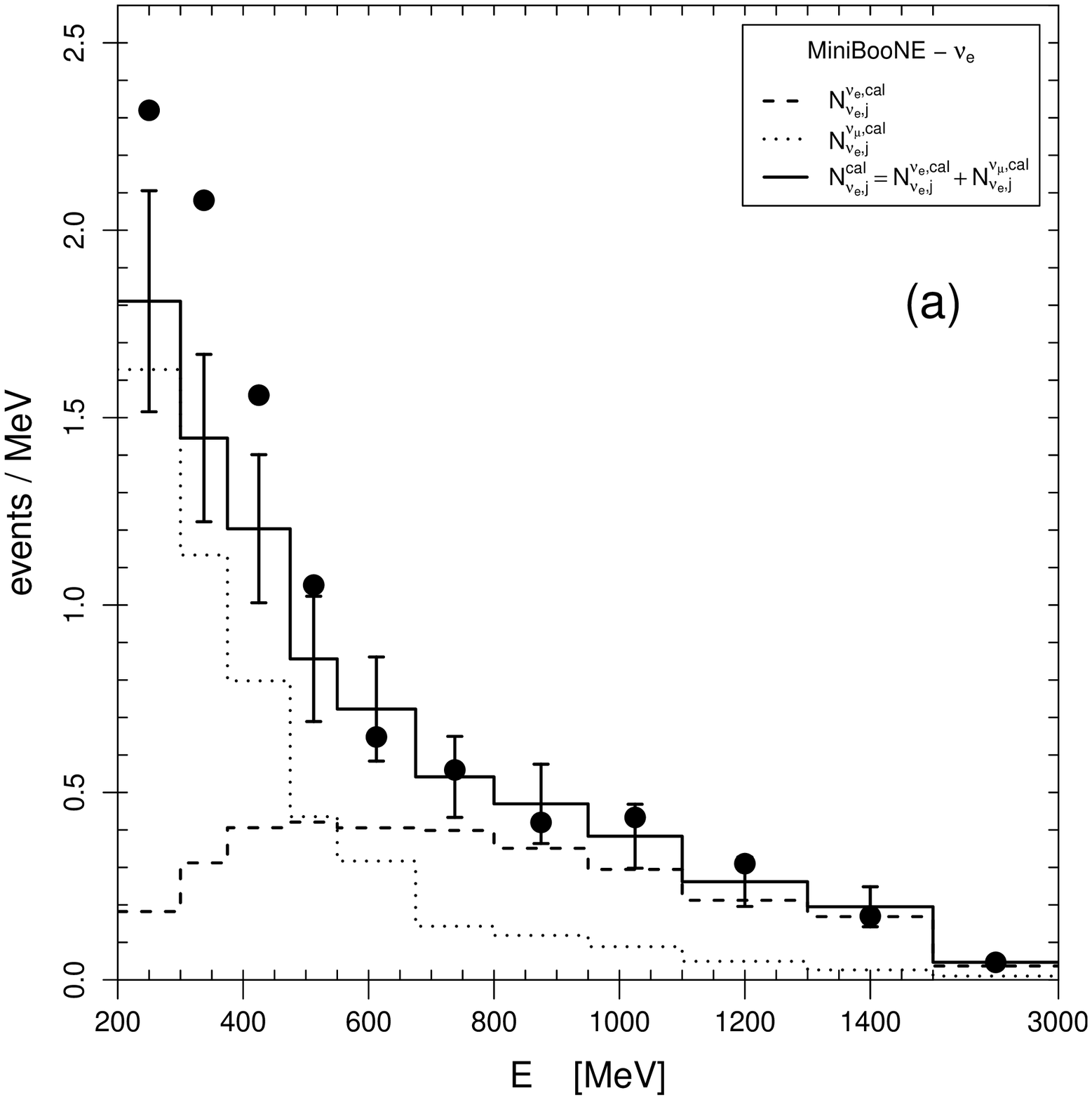}
&
\includegraphics*[bb=33 147 578 695, width=0.48\linewidth]{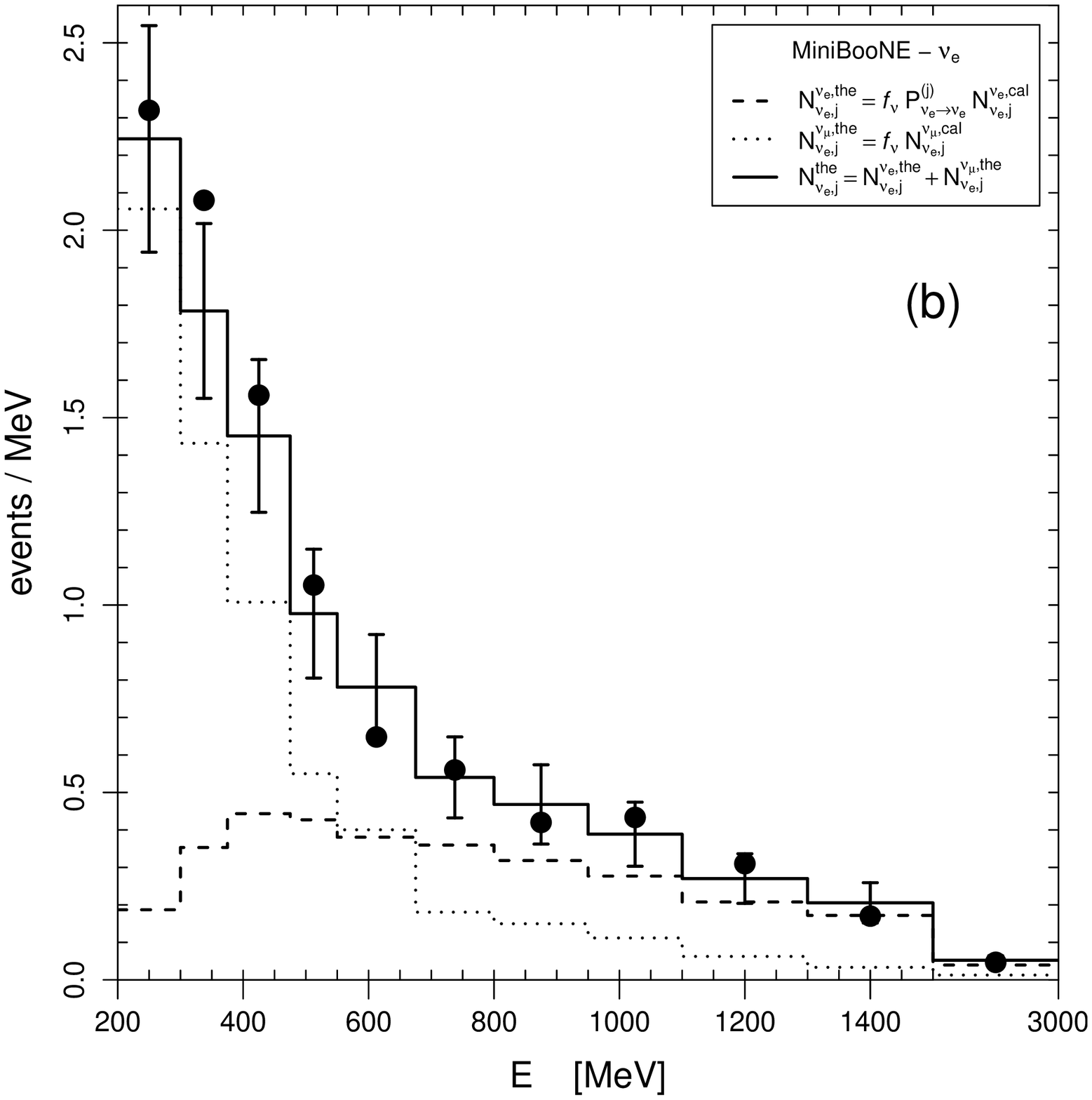}
\end{tabular}
\end{center}
\caption{ \label{008}
Expected number of $\nu_{e}$ events compared with MiniBooNE data,
represented by the black points.
The energy bins are numbered with the index $j$.
The uncertainty is represented by the vertical error bars,
which represent the sum of statistical and uncorrelated systematic uncertainties.
(a)
Expected number of $\nu_{e}$-like events
$N_{\nu_{e},j}^{\text{cal}}$
calculated by the MiniBooNE collaboration.
$N_{\nu_{e},j}^{\text{cal}}$
is given by the sum of
the $\nu_{e}$-induced events ($N_{\nu_{e},j}^{\nu_{e},\text{cal}}$)
and
the misidentified $\nu_{\mu}$-induced events ($N_{\nu_{e},j}^{\nu_{\mu},\text{cal}}$).
(b)
Best-fit value of the number of $\nu_{e}$-like events
$N_{\nu_{e},j}^{\text{the}}$
obtained with the hypothesis of $\nu_{e}$ disappearance.
$N_{\nu_{e},j}^{\text{the}}$
is given by the sum of
$ N_{\nu_{e},j}^{\nu_{e},\text{the}} = f_{\nu} P_{\nu_{e}\to\nu_{e}}^{(j)} N_{\nu_{e},j}^{\nu_{e},\text{cal}} $
and
$ N_{\nu_{e},j}^{\nu_{\mu},\text{the}} = f_{\nu} N_{\nu_{\mu}}^{\text{cal}} $.
The best-fit values of $f_{\nu}$, $\sin^2 2\vartheta$ and $\Delta{m}^2$
are those in the first column of Tab.~\ref{017}
(MB$\nu$).
}
\end{figure*}

\section{\label{009}MiniBooNE}

The MiniBooNE experiment was made with the purpose of checking the indication of
$\bar\nu_{\mu}\to\bar\nu_{e}$ oscillations generated by a $ \Delta{m}^2 \gtrsim 0.1 \, \text{eV}^2 $
found in the LSND experiment \cite{hep-ex/0104049}.
The MiniBooNE collaboration did not find any indication of such oscillations in the
$\nu_{\mu}\to\nu_{e}$ channel \cite{0704.1500,0812.2243}.
On the other hand,
the MiniBooNE collaboration found an anomalous excess of low-energy $\nu_{e}$-like events
in the data on the search for $\nu_{\mu}\to\nu_{e}$ oscillations \cite{0704.1500,0812.2243},
as shown in Fig.~\ref{008}a.

As in Refs.~\cite{0707.4593,0902.1992},
we consider an explanation of the low-energy MiniBooNE anomaly
based on the possible short-baseline disappearance of electron neutrinos,
taking into account a possible overall normalization factor $f_{\nu}$
of the calculated
$\nu_{e}$-induced and misidentified $\nu_{\mu}$-induced events
which contribute to the observed number of $\nu_{e}$-like events.
The normalization factor $f_{\nu}$ could be due mainly to the uncertainty
of the calculated neutrino flux (see Ref.\cite{physics/0609129}).
Since the misidentified $\nu_{\mu}$-induced and $\nu_{e}$-induced events dominate, respectively, at low and high energies
(see Fig.~\ref{008}a),
the low-energy excess can be fitted with $f_{\nu}>1$
and the high-energy data can be fitted compensating $f_{\nu}>1$ with the disappearance of $\nu_{e}$'s.

In Refs.~\cite{0707.4593,0902.1992} we considered
only very-short-baseline $\nu_{e}$ disappearance due to a $ \Delta{m}^2 \gtrsim 20 \, \text{eV}^2 $,
which generates a survival probability $P_{\nu_{e}\to\nu_{e}}$ which is constant in the MiniBooNE energy range,
from 200 to 3000 MeV.
In this paper we extend the analysis to lower values of $ \Delta{m}^2 $,
considering the resulting energy dependence of the survival probability.
In this case,
the theoretical number of $\nu_{e}$-like events in the $j$th energy bin is given by
\begin{equation}
N_{\nu_{e},j}^{\text{the}}
=
N_{\nu_{e},j}^{\nu_{e},\text{the}}
+
N_{\nu_{e},j}^{\nu_{\mu},\text{the}}
\,,
\label{010}
\end{equation}
where
\begin{equation}
N_{\nu_{e},j}^{\nu_{e},\text{the}} = f_{\nu} P_{\nu_{e}\to\nu_{e}}^{(j)} N_{\nu_{e},j}^{\nu_{e},\text{cal}}
\label{011}
\end{equation}
is the number of $\nu_{e}$-induced events
and
\begin{equation}
N_{\nu_{e},j}^{\nu_{\mu},\text{the}} = f_{\nu} N_{\nu_{e},j}^{\nu_{\mu},\text{cal}}
\label{012}
\end{equation}
is the number of misidentified $\nu_{\mu}$-induced events.
Here
$N_{\nu_{e},j}^{\nu_{e},\text{cal}}$
and
$N_{\nu_{e},j}^{\nu_{\mu},\text{cal}}$
are, respectively,
the number of $\nu_{e}$-induced and misidentified $\nu_{\mu}$-induced events
calculated by the MiniBooNE collaboration for the $j$th energy bin
\cite{AguilarArevalo:2008rc-dr,Louis-2009}.
$P_{\nu_{e}\to\nu_{e}}^{(j)}$ is the survival probability of electron neutrinos in Eq.~(\ref{004}) averaged
in the $j$th energy bin.
The average in each bin is calculated using the ntuple-file of 17,037 predicted muon-to-electron neutrino full transmutation events
given in Ref.~\cite{AguilarArevalo:2008rc-dr},
which contains information on reconstructed neutrino energy, true neutrino energy, neutrino baseline and event weight for each event.

The MiniBooNE measurement of a ratio $ 1.21 \pm 0.24 $
of detected and predicted charged-current quasi-elastic $\nu_{\mu}$ events \cite{0706.0926}
allows a value of $f_{\nu}$ as large as about 15\%.
In Ref.~\cite{0902.1992} we used this estimate of the uncertainty of $f_{\nu}$
in order to constrain its value in the least-squares analysis.
Here we use directly the $\nu_{\mu}$ data given in Ref.~\cite{AguilarArevalo:2008rc-dr}
for the construction of the MiniBooNE least-squares function
\begin{align}
\null & \null
\chi^2_{\text{MB$\nu$}}
=
\sum_{j=1}^{11}
\left(
\frac
{ N_{\nu_{e},j}^{\text{the}} - N_{\nu_{e},j}^{\text{exp}} }
{ \sigma_{\nu_{e},j} }
\right)^2
\nonumber
\\
\null & \null
+
\sum_{j,k=1}^{8}
\left( N_{\nu_{\mu},j}^{\text{the}} - N_{\nu_{\mu},j}^{\text{exp}} \right)
(V_{\nu_{\mu}}^{-1})_{jk}
\left( N_{\nu_{\mu},k}^{\text{the}} - N_{\nu_{\mu},k}^{\text{exp}} \right)
\,.
\label{013}
\end{align}
Here
$N_{\nu_{e},j}^{\text{exp}}$ are the numbers of measured $\nu_{e}$-like events in 11 reconstructed neutrino energy bins
and
$N_{\nu_{\mu},j}^{\text{exp}}$ are the numbers of measured $\nu_{\mu}$ charged-current quasi-elastic events in 8 reconstructed neutrino energy bins.
The theoretical number of $\nu_{\mu}$ events in the $j$th energy bin is given by
\begin{equation}
N_{\nu_{\mu},j}^{\text{the}} = f_{\nu} N_{\nu_{\mu},j}^{\text{cal}}
\,,
\label{014}
\end{equation}
where
$N_{\nu_{\mu},j}^{\text{cal}}$
is the number of $\nu_{\mu}$ events calculated by the MiniBooNE collaboration
\cite{AguilarArevalo:2008rc-dr}.
In order to take into account the correct statistical uncertainty corresponding to the rescaling of
the number of $\nu_{\mu}$ events due to $f_{\nu}$ in Eq.~(\ref{014}),
we used the covariance matrix $V_{\nu_{\mu}}$ given by
\begin{equation}
(V_{\nu_{\mu}})_{jk}
=
(V_{\nu_{\mu}}^{\text{cal}})_{jk} + \left( f_{\nu} - 1 \right) N_{\nu_{\mu},j}^{\text{cal}} \delta_{jk}
\,,
\label{015}
\end{equation}
where $V_{\nu_{\mu}}^{\text{cal}}$ is the $8\times8$ covariance matrix of $\nu_{\mu}$ events
presented by the MiniBooNE collaboration in Ref.~\cite{AguilarArevalo:2008rc-dr}.
We did not use the complete $19\times19$ covariance matrix of $\nu_{e}$ and $\nu_{\mu}$ events
given in Ref.~\cite{AguilarArevalo:2008rc-dr} because the correlations involving $\nu_{e}$ events
have been obtained without taking into account the energy-dependent disappearance of electron neutrinos
that we want to test.
Assuming the correlations given in that $19\times19$ covariance matrix would suppress the energy dependence of
$P_{\nu_{e}\to\nu_{e}}^{(j)}$.
Therefore,
for the uncertainties $\sigma_{\nu_{e},j}$ in Eq.~(\ref{013})
we used only the diagonal elements of the $\nu_{e}$ covariance matrix $V_{\nu_{e}}^{\text{cal}}$
given in Ref.~\cite{AguilarArevalo:2008rc-dr},
corrected by the change of statistical uncertainty corresponding to the variation of expected events due to $f_{\nu}$ and $P_{\nu_{e}\to\nu_{e}}^{(j)}$
in Eqs.~(\ref{010})--(\ref{012}):
\begin{equation}
\sigma_{\nu_{e},j}^2
=
(V_{\nu_{e}}^{\text{cal}})_{jj}
+
N_{\nu_{e},j}^{\text{the}}
-
N_{\nu_{e},j}^{\text{cal}}
\,,
\label{016}
\end{equation}
with
$ N_{\nu_{e},j}^{\text{cal}} = N_{\nu_{e},j}^{\nu_{e},\text{cal}} + N_{\nu_{e},j}^{\nu_{\mu},\text{cal}} $.

\begin{table*}
\begin{center}
\begin{tabular}{ccccccc}
&
&
MB$\nu$
&
Ga
&
MB$\nu$+Ga
&
Re+$^3$H
&
(MB$\nu$+Ga)+(Re+$^3$H)
\\
\hline
 Null Hyp. & $\chi^{2}$ & $ 14.3+5.4 $ & $ 9.4 $ & $ $ & $ 51.5 $ & $ $ \\
 & NDF & $ 3+16 $ & $ 4 $ & $ $ & $ 58 $ & $ $ \\
 & GoF & $ 0.41 $ & $ 0.051 $ & $ $ & $ 0.71 $ & $ $ \\
\hline Our Hyp. & $\chi^{2}_{\text{min}}$ & $ 2.0+7.6 $ & $ 1.8 $ & $ 2.2+9.2 $ & $ 49.1 $ & $ 4.1+63.4 $ \\
 & NDF & $ 16 $ & $ 2 $ & $ 20 $ & $ 56 $ & $ 78 $ \\
 & GoF & $ 0.89 $ & $ 0.40 $ & $ 0.93 $ & $ 0.73 $ & $ 0.80 $ \\
 & $\sin^22\vartheta_{\text{bf}}$ & $ 0.32 $ & $ 0.27 $ & $ 0.28 $ & $ 0.042 $ & $ 0.062 $ \\
 & $\Delta{m}^2_{\text{bf}}$ & $ 1.84 $ & $ 2.09 $ & $ 1.92 $ & $ 1.85 $ & $ 1.85 $ \\
 & $f_{\nu}^{\text{bf}}$ & $ 1.26 $ & $ $ & $ 1.25 $ & $ $ & $ 1.17 $ \\
\hline PG & $\Delta\chi^{2}_{\text{min}}$ & $ $ & $ $ & $ 0.098 $ & $ 0.01 $ & $ 6.97 $ \\
 & NDF & $ $ & $ $ & $ 2 $ & $ 2 $ & $ 2 $ \\
 & GoF & $ $ & $ $ & $ 0.95 $ & $ 0.99 $ & $ 0.03 $ \\
\hline
\end{tabular}
\caption{ \label{017}
Values of
$\chi^{2}$,
number of degrees of freedom (NDF) and
goodness-of-fit (GoF)
for the fit of different combinations of
MiniBooNE (MB$\nu$),
Gallium (Ga),
and
reactor (Re) data.
The first three lines correspond to the case of $f_{\nu}=1$ and no oscillations (Null Hyp.).
The following six lines correspond to the case $f_{\nu}>1$ and $\nu_{e}$ disappearance (Our Hyp.).
The last three lines give the parameter goodness-of-fit (PG) \protect\cite{hep-ph/0304176}.
In the MB$\nu$ column,
the value of $\chi^{2}$ in the Null. Hyp.,
the number of degrees of freedom in the Null. Hyp.
(which is equal to the number of energy bins),
and the value of $\chi^{2}_{\text{min}}$ in Our Hyp.
are shown as the sum of the
contributions of the first three low-energy $\nu_{e}$ bins and
the other $\nu_{e}$ and $\nu_{\mu}$ energy bins.
In the MB$\nu$+Ga and (MB$\nu$+Ga)+(Re+$^3$H) columns
the value of $\chi^{2}_{\text{min}}$ in Our Hyp.
is shown as the sum of the
contribution of the first three MiniBooNE low-energy $\nu_{e}$ bins and
the other contributions.
}
\end{center}
\end{table*}

The result of the minimization of $\chi^2_{\text{MB$\nu$}}$ is shown in Fig.~\ref{008}b,
in which the solid histogram corresponds to the best-fit values of
$f_{\nu}$,
$\sin^2 2\vartheta$ and
$\Delta{m}^2$
in the first column of Tab.~\ref{017}.
From Fig.~\ref{008}b, one can see that the fit is acceptable for all the $\nu_{e}$ energy bins,
including the first three bins which are out-of-fit in Fig.~\ref{008}a.
In Tab.~\ref{017} we give separately the contribution to
$\chi^2_{\text{MB$\nu$}}$ of the first three low-energy $\nu_{e}$ bins
and the sum of the contributions of the other $\nu_{e}$ energy bins and all the $\nu_{\mu}$ energy bins.
In this way one can see that with $f_{\nu}=1$ and $P_{\nu_{e}\to\nu_{e}}=1$ (Null Hypothesis),
although the global value $\chi^2=19.7$
is compatible with the number of degrees of freedom,
$\text{NDF}=19$,
almost all the $\chi^2$ is due to the
anomalous contribution $14.3$ of the first three low-energy $\nu_{e}$ bins,
whereas the other 16 $\nu_{e}$ and $\nu_{\mu}$ energy bins are overfitted,
with the excessively small $\chi^2$ contribution of $5.4$.
This overfitting, which is probably due to an overestimate of the uncertainties,
remains in the fit of the data with our hypothesis of $f_{\nu}>1$ and $\nu_{e}$ disappearance.
On the other hand,
our hypothesis clearly explains the low-energy anomaly reducing the $\chi^2$ contribution of the
first three low-energy $\nu_{e}$ bins to the acceptable best-fit value of $2.0$.

\begin{figure}[t!]
\begin{center}
\includegraphics*[bb=23 144 572 704, width=\linewidth]{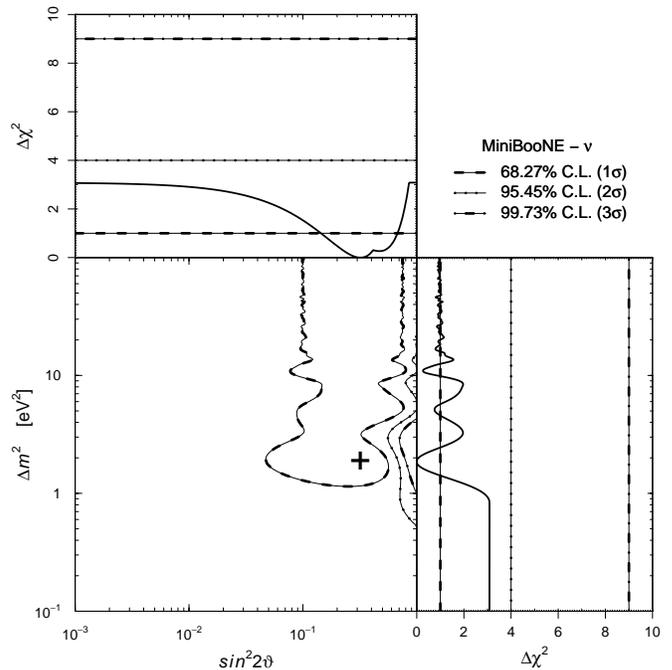}
\end{center}
\caption{ \label{018}
Allowed regions in the
$\sin^{2}2\vartheta$--$\Delta{m}^{2}$ plane
and
marginal $\Delta\chi^{2}$'s
for
$\sin^{2}2\vartheta$ and $\Delta{m}^{2}$
obtained from the fit of MiniBooNE neutrino data.
The best-fit point is indicated by a cross.
}
\end{figure}

Figure~\ref{018} shows the allowed regions
in the
$\sin^{2}2\vartheta$--$\Delta{m}^{2}$ plane
and
the
marginal $\Delta\chi^{2} = \chi^2 - \chi^2_{\text{min}}$'s
for
$\sin^{2}2\vartheta$ and $\Delta{m}^{2}$,
from which one can infer the corresponding uncorrelated allowed intervals.
One can see that the indication in favor of neutrino oscillations is not strong,
being at the level of about 79\% C.L. ($1.2\sigma$).
It is interesting to notice that the best-fit value of $\Delta{m}^2$ is about $2\,\text{eV}^2$,
which is approximately the same best-fit value obtained in
Ref.~\cite{0711.4222}
from the fit of
the neutrino data of Gallium radioactive source experiments
and
the antineutrino data of the Bugey and Chooz reactor experiments
under the hypothesis of $\nu_{e}$ and $\bar\nu_{e}$ disappearance.
The results of the combined analysis of MiniBooNE neutrino data
and the data of these other experiments is discussed in the following Sections.

\section{\label{019}Gallium Radioactive Source Experiments}

The GALLEX
\cite{Anselmann:1995ar,Hampel:1998fc,1001.2731}
and
SAGE
\cite{Abdurashitov:1996dp,hep-ph/9803418,nucl-ex/0512041,0901.2200}
collaborations
tested the respective Gallium solar neutrino detectors
in so-called
"Gallium radioactive source experiments"
which consist in the detection of electron neutrinos
produced by intense artificial ${}^{51}\text{Cr}$ and ${}^{37}\text{Ar}$ radioactive sources
placed inside the detectors.
Taking into account the uncertainty of the cross section of
the detection process
$ \nu_{e} + {}^{71}\text{Ga} \to {}^{71}\text{Ge} + e^{-} $
estimated in Ref.~\cite{hep-ph/9710491},
the ratio $R$ of measured and predicted ${}^{71}\text{Ge}$ event rates are
\begin{align}
R^{\text{GALLEX}}_{\text{Cr1}}
=
\null & \null
0.95
{}^{+0.11}_{-0.12}
\,,
\label{020}
\\
R^{\text{GALLEX}}_{\text{Cr2}}
=
\null & \null
0.81
{}^{+0.10}_{-0.11}
\,,
\label{021}
\\
R^{\text{SAGE}}_{\text{Cr}}
=
\null & \null
0.95
{}^{+0.12}_{-0.12}
\,,
\label{022}
\\
R^{\text{SAGE}}_{\text{Ar}}
=
\null & \null
0.79
{}^{+0.09}_{-0.10}
\,,
\label{023}
\end{align}
and the average ratio is
\begin{equation}
R^{\text{Ga}}
=
0.86
{}^{+0.05}_{-0.05}
\,.
\label{024}
\end{equation}
Thus,
the number of measured events is about $2.7\sigma$ smaller than
the prediction.

The theoretical prediction of the rate is based on the calculation of the detection cross section
presented in Ref.~\cite{hep-ph/9710491}.
It is possible that a part of the observed deficit is due to an overestimation of this cross section
\cite{nucl-ex/0512041,hep-ph/0605186,0901.2200},
because only the cross section of the transition
from the ground state of ${}^{71}\text{Ga}$ to the ground state of ${}^{71}\text{Ge}$
is known with precision from the measured rate of electron capture decay of
${}^{71}\text{Ge}$ to ${}^{71}\text{Ga}$.
Electron neutrinos produced by ${}^{51}\text{Cr}$ and ${}^{37}\text{Ar}$ radioactive sources
can be absorbed also through transitions from the ground state of ${}^{71}\text{Ga}$
to two excited states of ${}^{71}\text{Ge}$,
with cross sections which are inferred using a nuclear model
from $ p + {}^{71}\text{Ga} \to {}^{71}\text{Ge} + n $ measurements
\cite{Krofcheck:1985fg}.
This calculation has large uncertainties
\cite{nucl-th/9503017,nucl-th/9804011}.
However,
since the contribution of the transitions to the two excited states is only 5\%
\cite{hep-ph/9710491},
even the complete absence of such transitions
would reduce $R^{\text{Ga}}$ to about
$
0.90
{}^{+0.05}_{-0.05}
$,
leaving an anomaly of about $1.8\sigma$.

Here we consider the electron neutrino disappearance explanation of the
Gallium radioactive source experiments anomaly
\cite{hep-ph/9411414,Laveder:2007zz,hep-ph/0610352,0707.4593,0711.4222,0902.1992}
(another interesting explanation through quantum decoherence in neutrino oscillations
has been proposed in Ref.~\cite{0805.2098}).

\begin{figure}[t!]
\begin{center}
\includegraphics*[bb=23 144 572 704, width=\linewidth]{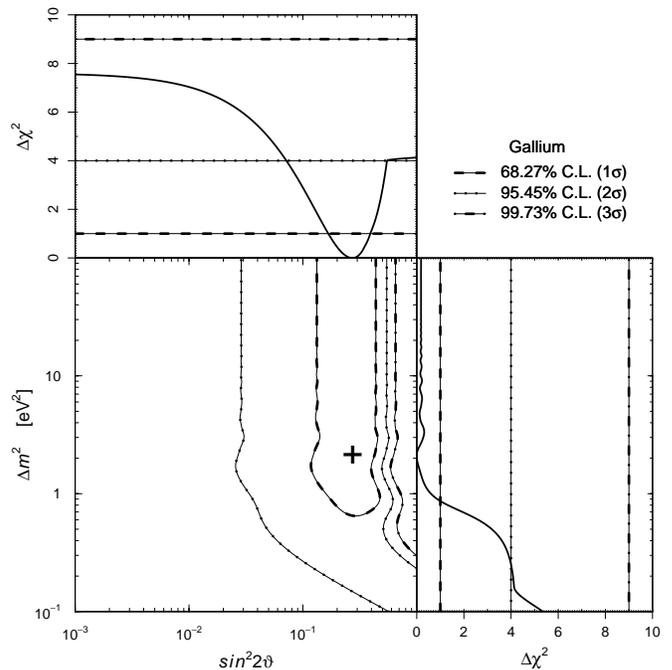}
\end{center}
\caption{ \label{025}
Allowed regions in the
$\sin^{2}2\vartheta$--$\Delta{m}^{2}$ plane
and
marginal $\Delta\chi^{2}$'s
for
$\sin^{2}2\vartheta$ and $\Delta{m}^{2}$
obtained from the
combined fit of the results of
the two GALLEX ${}^{51}\text{Cr}$ radioactive source experiments
and
the SAGE
${}^{51}\text{Cr}$ and ${}^{37}\text{Ar}$ radioactive source experiments.
The best-fit point corresponding to $\chi^2_{\text{min}}$ is indicated by a cross.
}
\end{figure}

In Ref.~\cite{0711.4222} we have analyzed the data
of the Gallium radioactive source experiments
in terms of the effective survival probability in Eq.~(\ref{004}).
Here we update that analysis taking into account the revised value of
$R^{\text{GALLEX}}_{\text{Cr1}}$ in Eq.~(\ref{020})
published recently in Ref.~\cite{1001.2731}
and taking into account the asymmetric uncertainties of
$R^{\text{GALLEX}}_{\text{Cr1}}$, $R^{\text{GALLEX}}_{\text{Cr2}}$ and $R^{\text{SAGE}}_{\text{Ar}}$
(which have been symmetrized for simplicity in the analysis presented in Ref.~\cite{0711.4222}).
Following the method described in Ref.~\cite{0711.4222},
we obtained the best-fit values of
$\sin^2 2\vartheta$ and $\Delta{m}^2$
in the second column of Tab.~\ref{017}
and the allowed regions
in the
$\sin^{2}2\vartheta$--$\Delta{m}^{2}$ plane
shown in Fig.~\ref{025}.
The indication in favor of neutrino oscillations is
at the level of about 98\% C.L. ($2.3\sigma$).

From Tab.~\ref{018} and the comparison of Figs.~\ref{018} and \ref{025}
one can see that the fits of MiniBooNE and Gallium data
lead to remarkably similar results:
the best-fit values of the oscillation parameters are very close
and the allowed regions
in the
$\sin^{2}2\vartheta$--$\Delta{m}^{2}$ plane
are highly compatible.
This is certainly an impressive success of our hypothesis
of electron neutrino disappearance.

\begin{figure}[t!]
\begin{center}
\includegraphics*[bb=23 144 572 704, width=\linewidth]{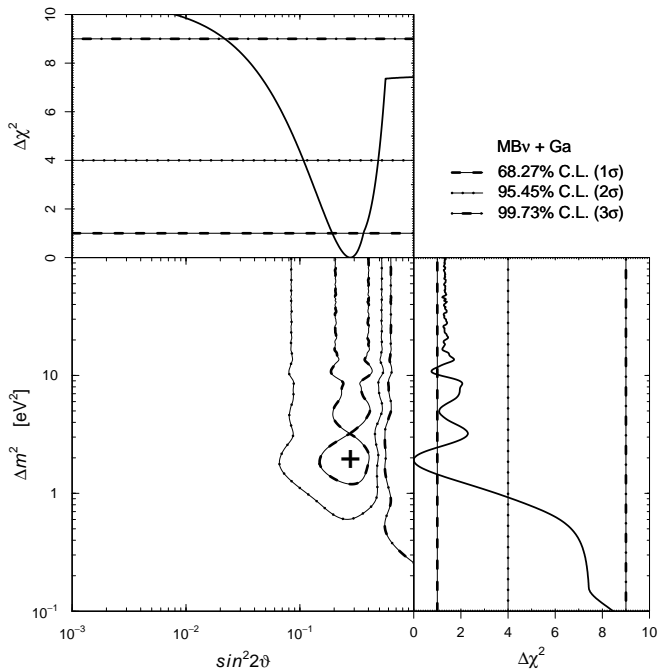}
\end{center}
\caption{ \label{026}
Allowed regions in the
$\sin^{2}2\vartheta$--$\Delta{m}^{2}$ plane
and
marginal $\Delta\chi^{2}$'s
for
$\sin^{2}2\vartheta$ and $\Delta{m}^{2}$
obtained from the
combined fit of the results of
MiniBooNE neutrino data
and the data of the gallium radioactive source experiments.
The best-fit point corresponding to $\chi^2_{\text{min}}$ is indicated by a cross.
}
\end{figure}

The results of the combined fit of MiniBooNE and Gallium data
are shown in the third column of Tab.~\ref{017}
and
in Fig.~\ref{026}.
The separate data sets are well fitted by the electron neutrino disappearance hypothesis:
the $\chi^2$ contribution of the first three MiniBooNE low-energy $\nu_{e}$ bins is
2.2,
that of the other 16 MiniBooNE $\nu_{e}$ and $\nu_{\mu}$ energy bins is
7.4,
and that of the 4 Gallium data is
1.9.
The consistency of the combined fit is also supported by the excellent
value of the parameter goodness-of-fit.
Combining the two data sets improves the indication in favor of neutrino oscillations to
the level of about 99.5\% C.L. ($2.8\sigma$).

\section{\label{027}Reactor and Tritium Experiments}

The indication of electron neutrino disappearance
that we have found from the analysis of MiniBooNE and Gallium data
must be confronted with the results of reactor electron antineutrino experiments.
Assuming CPT invariance, the survival probabilities of
neutrinos and antineutrinos are equal
(see Ref.~\cite{Giunti-Kim-2007}).
Thus, we can combine directly the results presented in the previous Section
with the results of the analysis of the data of the Bugey and Chooz reactor experiments
obtained in Ref.~\cite{0711.4222}.
We are encouraged in this task by the coincidence
of the best-fit value of $\Delta{m}^2$ at about $2\,\text{eV}^2$.

\begin{figure}[t!]
\begin{center}
\includegraphics*[bb=24 147 567 702, width=\linewidth]{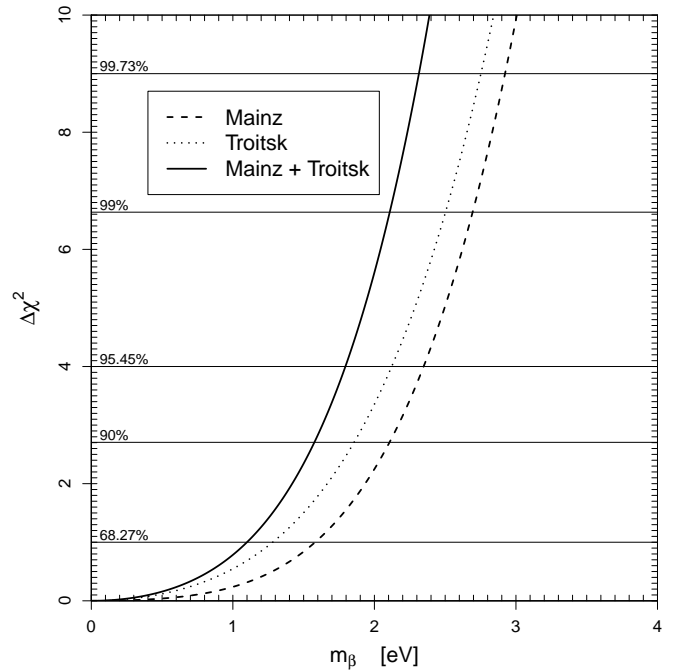}
\end{center}
\caption{ \label{035}
$\Delta\chi^2$
as a function of
$m_{\beta}$.
The horizontal lines correspond to the indicated value of confidence level.
The dashed and dotted lines have been
obtained, respectively, from the results in Eqs.~(\ref{030}) and (\ref{031})
of the Mainz and Troitsk
Tritium $\beta$-decay experiments.
The solid line is the result of the combined fit.
}
\end{figure}

In addition to reactor neutrino experiments,
also Tritium $\beta$-decay experiments give information on the
masses and mixing of neutrinos
through the measurement of the electron energy spectrum
in the process
\begin{equation}
{}^{3}\text{H}
\to
{}^{3}\text{He} + e^{-} + \bar\nu_{e}
\,.
\label{028}
\end{equation}
The most accurate measurements of the effective electron neutrino mass
(see Refs.~\cite{hep-ph/0211462,Giunti-Kim-2007})
\begin{equation}
m_{\beta}
=
\left(
\sum_{k} |U_{ek}|^2 m_{k}^2
\right)^{1/2}
\label{029}
\end{equation}
have been performed in the
Mainz \cite{hep-ex/0412056}
and
Troitsk \cite{Lobashev:2003kt}:
\begin{align}
m_{\beta}^2
=
\null & \null
-0.6 \pm 2.2 \pm 2.1 \, \text{eV}^2
\null & \null
\null & \null
(\text{Mainz})
\,,
\label{030}
\\
m_{\beta}^2
=
\null & \null
-2.3 \pm 2.5 \pm 2.0 \, \text{eV}^2
\null & \null
\null & \null
(\text{Troitsk})
\,.
\label{031}
\end{align}
These measurements can be interpreted and combined in order to derive upper bounds for
the effective mass $m_{\beta}$ through a $\chi^2$ analysis in the physical region
$m_{\beta}^2\geq0$.
In Fig.~\ref{035} we plotted the corresponding $\Delta\chi^2$'s
as a function of
$m_{\beta}$.
One can see that
\begin{align}
m_{\beta}
\leq
\null & \null
2.3 \, \text{eV}
\null & \null
\null & \null
(\text{Mainz, 95\% C.L.})
\,,
\label{032}
\\
m_{\beta}
\leq
\null & \null
2.0 \, \text{eV}
\null & \null
\null & \null
(\text{Troitsk, 95\% C.L.})
\,,
\label{033}
\end{align}
in approximate agreement with the corresponding values in Refs.~\cite{hep-ex/0412056,Lobashev:2003kt}.
The combined upper bound is
\begin{equation}
m_{\beta} \leq 1.8 \, \text{eV}
\qquad
(\text{Mainz+Troitsk, 95\% C.L.})
\,.
\label{034}
\end{equation}

\begin{figure}[t!]
\begin{center}
\includegraphics*[bb=23 144 572 704, width=\linewidth]{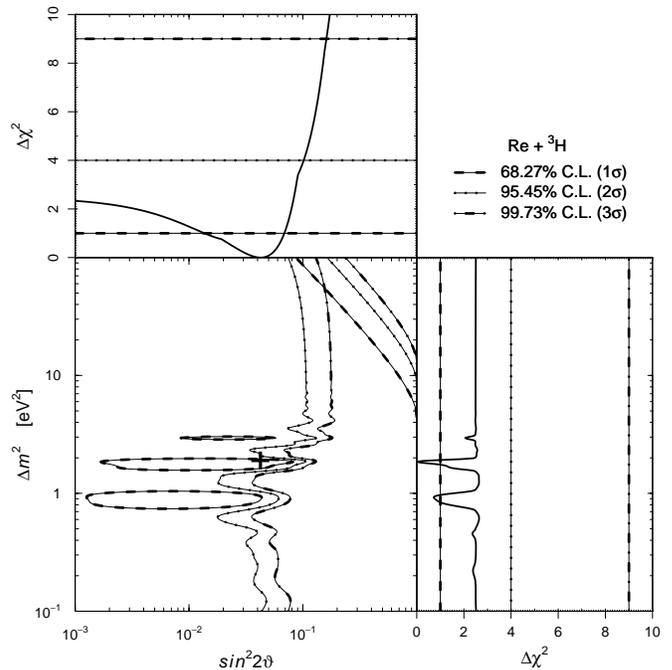}
\end{center}
\caption{ \label{040}
Allowed regions in the
$\sin^{2}2\vartheta$--$\Delta{m}^{2}$ plane
and
marginal $\Delta\chi^{2}$'s
for
$\sin^{2}2\vartheta$ and $\Delta{m}^{2}$
obtained from the combined fit of
the results of the Bugey and Chooz reactor experiments
and
the results of the Mainz and Troitsk Tritium $\beta$-decay experiments.
The three lines in the upper-right corner are the exclusion curves obtained from
the results of the Mainz and Troitsk
Tritium $\beta$-decay experiments alone.
The best-fit point corresponding to $\chi^2_{\text{min}}$ is indicated by a cross.
}
\end{figure}

In 3+1 four-neutrino schemes,
taking into account that the
mass splittings among
$m_{1}$,
$m_{2}$,
$m_{3}$
in Eqs.~(\ref{002}) and (\ref{003})
are negligible for a measurement of $m_{\beta}$ at the scale of 0.1 - 1 eV,
the effective mass is given by
\begin{equation}
m_{\beta}^2
\simeq
\left( 1 - |U_{e4}|^2 \right) m_{1}^2
+
|U_{e4}|^2 m_{4}^2
=
m_{1}^2 + |U_{e4}|^2 \Delta{m}^2_{41}
\,.
\label{036}
\end{equation}
In 3+1 schemes of the type in Eq.~(\ref{006}),
we have
\begin{equation}
m_{\beta}
\geq
|U_{e4}| \sqrt{\Delta{m}^2}
\,.
\label{037}
\end{equation}
The effective mixing angle in short-baseline electron neutrino disappearance experiments is
related to $|U_{e4}|$ by Eq.~(\ref{007}).
Inverting this relation and taking into account that
the value of
$|U_{e4}|$
must be small in order to fit the data of solar neutrino experiments
with neutrino oscillations,
we have
\begin{equation}
|U_{e4}|^2
=
\frac{1}{2}
\left(
1
-
\sqrt{ 1 - \sin^2 2\vartheta }
\right)
\,.
\label{038}
\end{equation}
In Fig.~\ref{040} we show the
limits in the $\sin^{2}2\vartheta$--$\Delta{m}^{2}$ plane
obtained from Eqs.~(\ref{037}) and (\ref{038})
and the results in Eqs.~(\ref{030}) and (\ref{031})
of the Mainz and Troitsk
Tritium $\beta$-decay experiments.
Notice that for small values of $\sin^2 2\vartheta$ the bounds are practically linear in the log-log plot in Fig.~\ref{040},
because in this case
$|U_{e4}|^2 \simeq \sin^2 2\vartheta / 4$
and
the inequality in Eq.~(\ref{037}) leads to
\begin{equation}
\log\Delta{m}^2
\lesssim
2 \log2
+
2 \log m_{\beta}^{\text{ub}}
-
\log\sin^2 2\vartheta
\,,
\label{039}
\end{equation}
where
$m_{\beta}^{\text{ub}}$ is the upper bound for $m_{\beta}$.

In the fourth column of Tab.~\ref{017}
and in Fig.~\ref{040}
we report the results of the analysis of the data of the Bugey and Chooz reactor experiments
presented in Ref.~\cite{0711.4222}\footnote{
As erratum,
let us notice that
in the fourth column of Table III in Ref.~\cite{0711.4222}
there is a small mistake in the evaluation of
the parameter goodness-of-fit.
The correct values are
$\Delta\chi^2_{\text{min}}=0.52$
and
$\text{GoF}=0.47$.
We also notice that in the version of
Ref.~\cite{0711.4222}
published in Phys. Rev. D
the value of $\sin^2 2\vartheta_{\text{bf}}$ for the Ga+Bu+Ch analysis
(last column of Table III in Ref.~\cite{0711.4222})
is different from the correct one, which is 0.054
(see the arXiv version of Ref.~\cite{0711.4222}).
},
with the addition in the analysis of the results of
the Mainz and Troitsk
Tritium $\beta$-decay experiments,
which affect the high-$\Delta{m}^2$ region.
As already commented in Ref.~\cite{0711.4222},
the reactor data are compatible with both the Null Hypothesis of absence of electron antineutrino disappearance
and Our Hypothesis of electron antineutrino disappearance,
with a hint in favor of electron antineutrino oscillations
due to a $\Delta{m}^2$ of about 2 eV.

\section{\label{041}Combined Analysis}

\begin{figure}[t!]
\begin{center}
\includegraphics*[bb=23 144 572 704, width=\linewidth]{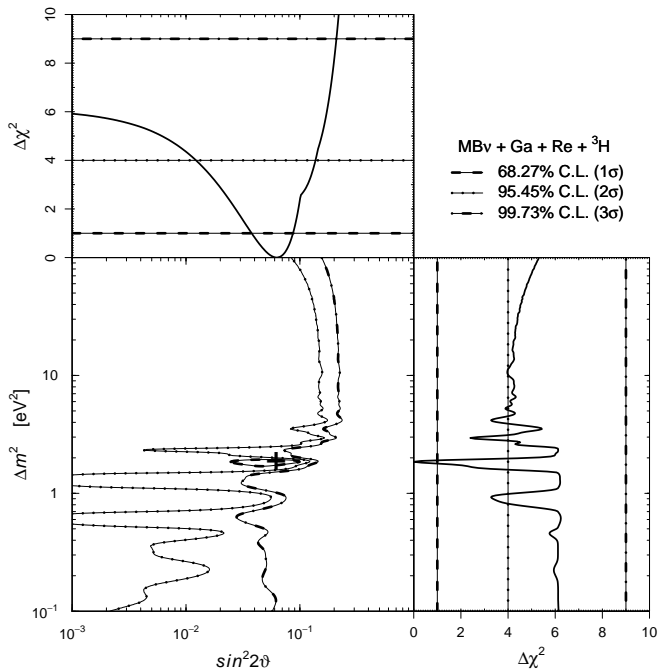}
\end{center}
\caption{ \label{042}
Allowed regions in the
$\sin^{2}2\vartheta$--$\Delta{m}^{2}$ plane
and
marginal $\Delta\chi^{2}$'s
for
$\sin^{2}2\vartheta$ and $\Delta{m}^{2}$
obtained from the
combined fit of the results of
MiniBooNE, Gallium, reactor and Tritium experiments.
The best-fit point corresponding to $\chi^2_{\text{min}}$ is indicated by a cross.
The three lines in the upper-right corner give the
$1\sigma$, $2\sigma$ and $3\sigma$
limits in the
$\sin^{2}2\vartheta$--$\Delta{m}^{2}$ plane
obtained from Eqs.~(\ref{029}) and (\ref{038})
and the results in Eqs.~(\ref{030}) and (\ref{031})
of the Mainz and Troitsk
Tritium $\beta$-decay experiments.
}
\end{figure}

The results of the combined analysis of MiniBooNE, Gallium, reactor and Tritium data
are presented in the last column of Tab.~\ref{017}
and in Fig.~\ref{042}.
One can see that the goodness-of-fit is high.
The separate data sets are fitted fairly well by the electron neutrino disappearance hypothesis:
the $\chi^2$ contribution of the first three MiniBooNE low-energy $\nu_{e}$ bins is
4.1,
that of the other 16 MiniBooNE $\nu_{e}$ and $\nu_{\mu}$ energy bins is
7.5,
that of the 4 Gallium data is
6.3,
that of the 56 reactor degrees of freedom is
49.0 and
that of the 2 Tritium degrees of freedom is
0.57.
On the other hand,
the
3\%
parameter goodness-of-fit of the combined analysis
of neutrino MiniBooNE and Gallium data
and antineutrino reactor and Tritium data
is rather low.

\begin{figure}[t!]
\begin{center}
\includegraphics*[bb=24 147 564 702, width=\linewidth]{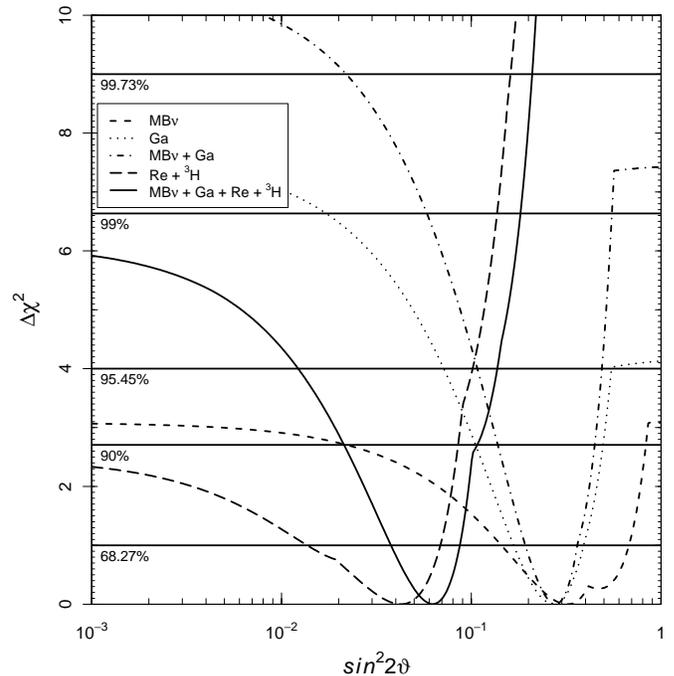}
\end{center}
\caption{ \label{043}
Marginal
$\Delta\chi^2 = \chi^2 - \chi^2_{\text{min}}$
for
$\sin^{2}2\vartheta$
obtained
from the analysis of different combinations of
MiniBooNE, Gallium, reactor and Tritium data.
The marginal
$\Delta\chi^2$
obtained from the analysis of reactor data alone
and that
obtained from the combined analysis of reactor and Tritium data
are shown by the same line, since they practically coincide.
}
\end{figure}

This low compatibility of the neutrino and antineutrino data sets
is illustrated in Fig.~\ref{043},
where we have plotted the marginal $\Delta\chi^{2}$'s
for
$\sin^{2}2\vartheta$
obtained with the analysis of different data sets\footnote{
We thank the anonymous referee of Phys. Rev. D for suggesting this interesting figure.}.
One can see that
\begin{enumerate}
\item
The neutrino MiniBooNE and Gallium data agree to indicate a value of
$\sin^{2}2\vartheta$
between about
0.11
and
0.48
at $2\sigma$.
\item
The antineutrino reactor data
indicate a value of
$\sin^{2}2\vartheta$
smaller than about
0.10
at $2\sigma$.
The Tritium data are practically irrelevant for the determination of $\sin^{2}2\vartheta$.
\item
The combined analysis is dominated by the reactor data and indicates a value of
$\sin^{2}2\vartheta$
between about
0.01
and
0.13
at $2\sigma$.
\end{enumerate}
The discrepancy between the
neutrino and antineutrino determinations of
$\sin^{2}2\vartheta$
is about $2\sigma$,
in rough agreement with the above-mentioned
3\%
parameter goodness-of-fit of the combined analysis.
In fact,
the $2\sigma$ disagreement between the neutrino and antineutrino data sets is entirely
due to the different requirements on the value of
$\sin^{2}2\vartheta$,
whereas they nicely agree on a best-fit value of $\Delta{m}^2$ at about $2\,\text{eV}^2$.

Since the 3\%
parameter goodness-of-fit of the combined analysis
shows a tension between the neutrino and antineutrino data
(under our $\nu_{e}$-disappearance hypothesis)
but is not sufficiently small to reject with confidence
the compatibility of the neutrino and antineutrino data sets\footnote{
For example,
the review on Statistics in the 2000 edition of the Review of Particle Physics
\cite{PDG-2000}
says that
if the goodness-of-fit
``is larger than an agreed-upon value (0.001,
0.01, or 0.05 are common choices), the data are consistent with the
assumptions''.
}, in the following part of this Section and in Section~\ref{045}
we consider the results and implications of the combined analysis.
In Section~\ref{058} we consider a possible
difference between the effective mixing angles in the
neutrino and antineutrino sectors.

Although the combined analysis of neutrino and antineutrino data favors
smaller values of
$\sin^{2}2\vartheta$
than those obtained from the analysis of MiniBooNE and Gallium data alone,
the fit of the MiniBooNE and Gallium data remains better than in the case of
no oscillations and $f_{\nu}=1$.

\begin{figure}[t!]
\begin{center}
\includegraphics*[bb=33 147 578 695, width=0.98\linewidth]{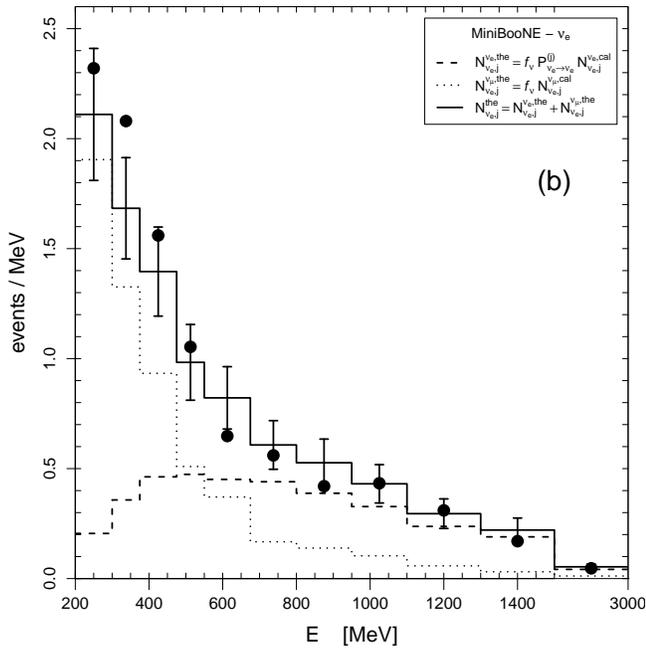}
\end{center}
\caption{ \label{044}
Expected number of MiniBooNE $\nu_{e}$ events
in the best-fit result of the combined analysis of MiniBooNE, Gallium, reactor and Tritium data
(last column in Tab.~\ref{017}).
The notation is the same as in Fig.~\ref{008}.
}
\end{figure}

Figure~\ref{044} shows the fit of MiniBooNE $\nu_{e}$ data corresponding to the
best-fit result of the combined analysis.
One can see that the fit of the first three low-energy bins
is not as good as that in Fig.~\ref{008}b,
but it is nevertheless acceptable and much better than that in Fig.~\ref{008}a.

For the Gallium source experiments,
the best-fit values of the oscillation parameters give
$
R^{\text{GALLEX}}_{\text{Cr1}}
=
R^{\text{GALLEX}}_{\text{Cr2}}
=
0.97
$,
$
R^{\text{SAGE}}_{\text{Cr}}
=
0.96
$
and
$
R^{\text{SAGE}}_{\text{Ar}}
=
0.96
$.
Therefore,
the experimental values of
$R^{\text{GALLEX}}_{\text{Cr1}}$
and
$R^{\text{SAGE}}_{\text{Cr}}$
in Eqs.~(\ref{020}) and (\ref{022}) are fitted very well
and
the Gallium $\chi^2$ contribution of
6.3
is almost equally due to the loose fits of
$R^{\text{GALLEX}}_{\text{Cr2}}$
and
$R^{\text{SAGE}}_{\text{Ar}}$
in Eqs.~(\ref{021}) and (\ref{023}).

Considering
the combined fit of the results of MiniBooNE, Gallium, reactor and Tritium data
as a fair indication in favor of a possible
short-baseline electron neutrino disappearance generated by the effective mixing parameters
$\Delta{m}^2 \simeq 2 \, \text{eV}$
and
$
0.01
\lesssim \sin^2 2\vartheta \lesssim
0.13
$,
in the next Section we present the corresponding predictions
for the effective neutrino masses in
$\beta$-decay and neutrinoless double-$\beta$-decay experiments
which could be measured in future experiments.

\section{\label{045}Predictions for Beta-Decay and Neutrinoless Double-Beta-Decay Experiments}

In this Section we present predictions
for the effective neutrino masses in
$\beta$-decay and neutrinoless double-$\beta$-decay experiments
obtained as a consequence of the combined fit of MiniBooNE, Gallium, reactor and Tritium data
discussed in the previous Section.

\begin{figure}[t!]
\begin{center}
\includegraphics*[bb=24 147 564 702, width=\linewidth]{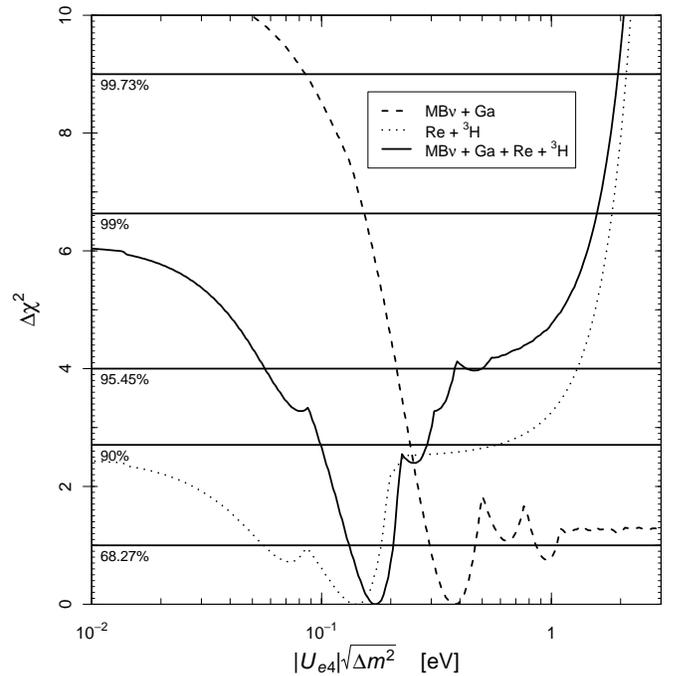}
\end{center}
\caption{ \label{049}
$\Delta\chi^2 = \chi^2 - \chi^2_{\text{min}}$
as a function of the contribution
$ |U_{e4}| \sqrt{\Delta{m}^2} $
to the effective $\beta$-decay electron-neutrino mass $m_{\beta}$
in four-neutrino schemes
obtained
from the analysis of
MiniBooNE and Gallium data (dashed line),
from the analysis of
reactor and Tritium data (dotted line),
and from the combined analysis of the two sets of data (solid line).
}
\end{figure}

Figure~\ref{049}
shows the residual $\Delta\chi^2 = \chi^2 - \chi^2_{\text{min}}$
as a function of the contribution
$|U_{e4}| \sqrt{\Delta{m}^2}$
to the effective mass $m_{\beta}$ in $\beta$-decay experiments
(see Eq.~(\ref{037})).
Since from the last column of Tab.~\ref{017} we have
$\sin^2 2\vartheta_{\text{bf}} \ll 1$,
we obtain
\begin{equation}
|U_{e4}|^2_{\text{bf}}
\simeq
\frac{\sin^2 2\vartheta_{\text{bf}}}{4}
=
0.016
\,,
\label{046}
\end{equation}
and the best-fit value of $|U_{e4}| \sqrt{\Delta{m}^2}$ is
\begin{equation}
\left( |U_{e4}| \sqrt{\Delta{m}^2} \right)_{\text{bf}}
=
0.17 \, \text{eV}
\,,
\label{047}
\end{equation}
and
\begin{equation}
0.06
\leq
|U_{e4}| \sqrt{\Delta{m}^2}
\leq
0.49
\, \text{eV}
\quad
\text{at $2\sigma$}
\,.
\label{048}
\end{equation}
This prediction is relevant for the KATRIN experiment \cite{0910.4862},
which is under construction and scheduled to start in 2012.
The expected sensitivity of about 0.2 eV at 90\% C.L. may be sufficient to observe a positive effect
if $|U_{e4}| \sqrt{\Delta{m}^2}$ is sufficiently large,
as allowed by $\Delta\chi^2$ in Fig.~\ref{049}.

If massive neutrinos are Majorana particles,
neutrinoless double-$\beta$ decay is possible,
with a decay rate proportional to the effective Majorana mass
(see Refs.~\cite{hep-ph/0202264,hep-ph/0211462,hep-ph/0405078,0708.1033,Giunti-Kim-2007})
\begin{equation}
m_{2\beta}
=
\left| \sum_{k} U_{ek}^2 m_{k} \right|
\,.
\label{050}
\end{equation}
The results of the combined fit of MiniBooNE, Gallium, reactor and Tritium data
discussed in Section~\ref{027} allow us to
estimate the contribution of the heaviest massive neutrino $\nu_{4}$
to $m_{2\beta}$, which is approximately given by $ |U_{e4}|^2 \sqrt{\Delta{m}^2} $,
taking into account the mass hierarchy in Eq.~(\ref{006}).

\begin{figure}[t!]
\begin{center}
\includegraphics*[bb=24 147 564 702, width=\linewidth]{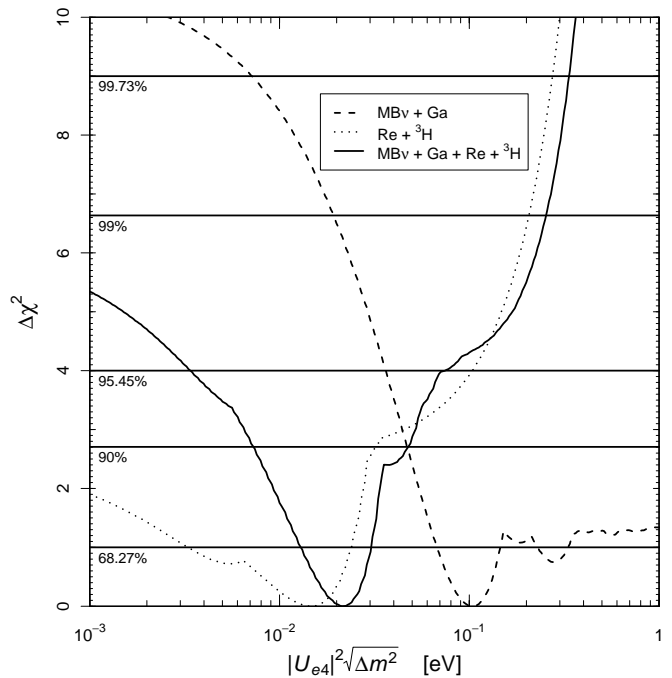}
\end{center}
\caption{ \label{057}
$\Delta\chi^2 = \chi^2 - \chi^2_{\text{min}}$
as a function of the contribution
$ |U_{e4}|^2 \sqrt{\Delta{m}^2} $
to the effective neutrinoless double-$\beta$ decay Majorana mass $m_{2\beta}$
in four-neutrino schemes
obtained
from the analysis of
MiniBooNE and Gallium data (dashed line),
from the analysis of
reactor and Tritium data (dotted line),
and from the combined analysis of the two sets of data (solid line).
}
\end{figure}

Figure~\ref{057}
shows $\Delta\chi^2 = \chi^2 - \chi^2_{\text{min}}$
as a function of the contribution
$ |U_{e4}|^2 \sqrt{\Delta{m}^2} $
in four-neutrino schemes to $m_{2\beta}$.
The best-fit value is:
\begin{equation}
\left( |U_{e4}|^2 \sqrt{\Delta{m}^2} \right)_{\text{bf}}
=
0.02 \, \text{eV}
\,,
\label{051}
\end{equation}
and
\begin{equation}
0.003
\leq
|U_{e4}|^2 \sqrt{\Delta{m}^2}
\leq
0.07
\, \text{eV}
\quad
\text{at $2\sigma$}
\,.
\label{052}
\end{equation}
This range must be confronted with the expected contributions to $m_{2\beta}$ coming from the three light massive neutrinos
$\nu_{1}$,
$\nu_{2}$,
$\nu_{3}$.
Assuming a hierarchy of masses,
\begin{equation}
m_{1} \ll m_{2} \ll m_{3} \ll m_{4}
\,,
\label{053}
\end{equation}
which is the most natural case compatible with the hierarchy in Eq.~(\ref{006}),
we have
\begin{equation}
m_{2\beta}
\simeq
\left|
U_{e2}^2 \sqrt{\Delta{m}^2_{\text{SOL}}}
+
U_{e3}^2 \sqrt{\Delta{m}^2_{\text{ATM}}}
+
U_{e4}^2 \sqrt{\Delta{m}^2}
\right|
\,,
\label{054}
\end{equation}
where we have neglected the contribution of the lightest massive neutrino $\nu_{1}$.
From the $3\sigma$ upper limits of the three-neutrino mixing parameters given in Ref.~\cite{hep-ph/0405172},
we obtain
\begin{equation}
2 \times 10^{-3}
\lesssim
|U_{e2}|^2 \sqrt{\Delta{m}^2_{\text{SOL}}}
\lesssim
4 \times 10^{-3} \, \text{eV}
\,,
\label{055}
\end{equation}
\begin{equation}
|U_{e3}|^2 \sqrt{\Delta{m}^2_{\text{ATM}}}
\lesssim
3 \times 10^{-3} \, \text{eV}
\,.
\label{056}
\end{equation}
Therefore,
strong cancellations between the contributions of $\nu_{2}$ and $\nu_{3}$ are possible
(albeit not likely \cite{hep-ph/9906275}),
whereas the range in Eq.~(\ref{052})
disfavors strong cancellations between the contributions of $\nu_{2}$ and $\nu_{3}$
and the contribution of $\nu_{4}$.
In this case,
$m_{2\beta} \simeq |U_{e4}|^2 \sqrt{\Delta{m}^2}$
leading to a possible observation of neutrinoless double-$\beta$ decay
in future experiments which will be sensitive to values of $m_{2\beta}$ smaller that $10^{-1} \, \text{eV}$
(e.g. CUORE \cite{Pedretti:2008zz}, EXO \cite{0909.1826}, SuperNEMO \cite{0909.3167}; see the review in Ref.~\cite{0708.1033}).

On the other hand,
if neutrinoless double-$\beta$ decay experiments which are sensitive to values of $m_{2\beta}$ of the order of $10^{-1} \, \text{eV}$
(e.g. CUORICINO \cite{0802.3439}, GERDA \cite{hep-ex/0404039}, Majorana \cite{0910.4598}; see the review in Ref.~\cite{0708.1033})
will see a positive signal,
maybe compatible with the signal asserted in Ref.~\cite{Klapdor-Kleingrothaus:2006ff},
the mass hierarchy in Eq.~(\ref{053})
will become unlikely and the favorite 3+1 four-neutrino schemes
will be those in which the three light neutrinos
$\nu_{1}$,
$\nu_{2}$ and
$\nu_{3}$
are almost degenerate at the mass scale of $m_{2\beta}$.

\section{\label{058}Mixing Angle Asymmetry?}

The tension between neutrino and antineutrino data
discussed in Section~\ref{041}
could be due to a difference of the effective mixing angles in the
neutrino and antineutrino sectors.
Such a difference could be due to
a violation of the fundamental CPT symmetry
or to another unknown mechanism.
Phenomenological analyses of different masses and mixings
for neutrinos and antineutrinos have been presented in several publications
\cite{hep-ph/0010178,hep-ph/0108199,hep-ph/0112226,hep-ph/0201080,hep-ph/0201134,hep-ph/0201211,hep-ph/0307127,hep-ph/0308299,hep-ph/0505133,hep-ph/0306226,0804.2820,0903.4318,0907.5487,0908.2993}.

In this section we consider the possibility
that neutrinos and antineutrinos have different effective masses and mixings
in short-baseline $\nu_{e}$ and $\bar\nu_{e}$ disappearance experiments.
We fit the neutrino and antineutrino data
with the survival probabilities
\begin{align}
P_{\nu_{e}\to\nu_{e}}^{\text{SBL}}(L,E)
=
\null & \null
1 - \sin^22\vartheta_{\nu} \sin^2 \left( \frac{\Delta{m}^{2}_{\nu} L}{4 E} \right)
\,,
\label{00222}
\\
P_{\bar\nu_{e}\to\bar\nu_{e}}^{\text{SBL}}(L,E)
=
\null & \null
1 - \sin^22\vartheta_{\bar\nu} \sin^2 \left( \frac{\Delta{m}^{2}_{\bar\nu} L}{4 E} \right)
\,.
\label{00333}
\end{align}
The results for the two fits are those presented in
Fig.~\ref{026} and the third column in Tab.~\ref{017} for neutrinos (MB$\nu$+Ga)
and
Fig.~\ref{040} and the fourth column in Tab.~\ref{017} for antineutrinos (Re+$^3$H).

\begin{figure}[t!]
\begin{center}
\includegraphics*[bb=24 147 564 702, width=\linewidth]{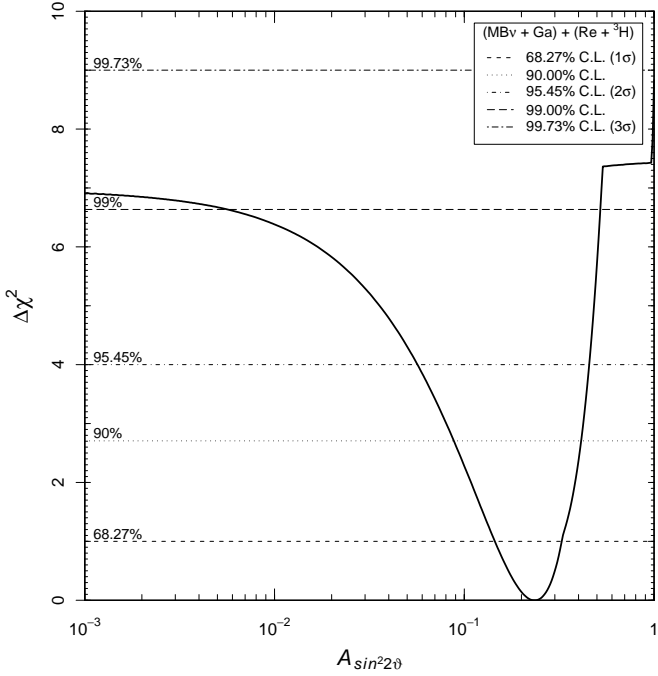}
\end{center}
\caption{ \label{059}
Marginal
$\Delta\chi^2 = \chi^2 - \chi^2_{\text{min}}$
as a function of the mixing angle asymmetry
$A_{\sin^22\vartheta}$.
}
\end{figure}

Since the fit of the data does not require a difference of
$\Delta{m}^{2}_{\nu}$ and $\Delta{m}^{2}_{\bar\nu}$,
we consider only the mixing angle asymmetry
\begin{equation}
A_{\sin^22\vartheta}
=
\sin^22\vartheta_{\nu} - \sin^22\vartheta_{\bar\nu}
\,.
\label{00334}
\end{equation}
Figure~\ref{059} shows the marginal $\Delta\chi^2$
as a function of $A_{\sin^22\vartheta}$.
The best-fit value of $A_{\sin^22\vartheta}$ is
\begin{equation}
A_{\sin^22\vartheta}^{\text{bf}}
=
0.23
\,,
\label{00335}
\end{equation}
and the $2\sigma$ allowed range of $A_{\sin^22\vartheta}$ is
\begin{equation}
0.06
\leq
A_{\sin^22\vartheta}
\leq
0.45
\,,
\label{00336}
\end{equation}
but there is no limit on the asymmetry at $3\sigma$.
The statistical significance of $A_{\sin^22\vartheta}>0$ is
99.14\% C.L.
($2.6\sigma$).

It is interesting to note that a difference between neutrino and antineutrino mixings
can be tested in $\beta$-decay experiments
by searching for different effective neutrino masses in
$\beta^{-}$ and $\beta^{+}$ decays.
The prediction for the contribution
$\left( |U_{e4}| \sqrt{\Delta{m}^2} \right)_{\bar\nu_{e}}$
to the effective electron antineutrino mass
in $\beta^{-}$ decays from the analysis of
antineutrino reactor and Tritium data can be obtained from the dotted line in Fig.~\ref{049}:
the best fit is
\begin{equation}
\left( |U_{e4}| \sqrt{\Delta{m}^2} \right)_{\bar\nu_{e}}^{\text{bf}}
=
0.14 \, \text{eV}
\,,
\label{bm1}
\end{equation}
and
\begin{equation}
\left( |U_{e4}| \sqrt{\Delta{m}^2} \right)_{\bar\nu_{e}}
\leq
1.29 \, \text{eV}
\quad
\text{at $2\sigma$}
\,.
\label{bm2}
\end{equation}
For the contribution
$\left( |U_{e4}| \sqrt{\Delta{m}^2} \right)_{\nu_{e}}$
to the effective electron neutrino mass
in $\beta^{+}$ decays we must consider the dashed line in Fig.~\ref{049},
which has been obtained from the analysis of MiniBooNE and Gallium neutrino data:
the best fit is
\begin{equation}
\left( |U_{e4}| \sqrt{\Delta{m}^2} \right)_{\nu_{e}}^{\text{bf}}
=
0.38 \, \text{eV}
\,,
\label{bp1}
\end{equation}
and
\begin{equation}
\left( |U_{e4}| \sqrt{\Delta{m}^2} \right)_{\nu_{e}}
\geq
0.21 \, \text{eV}
\quad
\text{at $2\sigma$}
\,.
\label{bp2}
\end{equation}
Unfortunately the existing and foreseen experiments are $\beta^{-}$-decay experiments
for which the contribution
$\left( |U_{e4}| \sqrt{\Delta{m}^2} \right)_{\bar\nu_{e}}$
to the effective electron antineutrino mass is expected to be small.
The future $\beta^{-}$-decay experiment will use either Tritium (KATRIN \cite{0910.4862})
or
${}^{187}\text{Re}$
(MARE \cite{hep-ex/0509038}).
If the mixing difference between the neutrino and antineutrino sectors
will be confirmed with highest confidence by future neutrino oscillation data
it will be interesting to study the possibility of making $\beta^{+}$-decay experiments
for the search of the effective electron antineutrino mass,
for which the dashed line in Fig.~\ref{049} and Eq.~(\ref{bp2})
give a reachable lower limit.

We do not consider here neutrinoless double-$\beta$ decay
in the case of a neutrino-antineutrino mixing difference,
since the Majorana nature of neutrinos requires a treatment
which goes well beyond the purposes of this paper
(see Ref.~\cite{hep-ph/0203261}).

\section{\label{060}Conclusions}

In this paper we have discussed a neutrino oscillation interpretation of the
MiniBooNE low-energy anomaly
and
the Gallium radioactive source experiments anomaly
in the framework of 3+1 four-neutrino mixing schemes.
We have shown that the combined fit of MiniBooNE and Gallium data
indicate a possible short-baseline electron neutrino disappearance
generated by effective oscillation parameters
$\Delta{m}^2 \gtrsim 0.1 \, \text{eV}^2$
and
$
0.11
\leq \sin^2 2\vartheta \leq
0.48
$
at $2\sigma$,
with best fit at
$\Delta{m}^2 \simeq 2 \, \text{eV}^2$
and
$\sin^2 2\vartheta \simeq 0.3$
(see Fig.~\ref{026}).

We have also considered the data of the Bugey and Chooz reactor neutrino oscillation experiments
and the results of the Mainz and Troitsk
Tritium $\beta$-decay experiments,
which imply an upper bound on the effective electron neutrino mass of about 2 eV
(see Fig.~\ref{035} and the combined upper bound in Eq.~(\ref{034})).
As already discussed in Ref.~\cite{0711.4222},
the Bugey data give a faint indication
of a possible short-baseline electron neutrino disappearance
generated by effective oscillation parameters
$\Delta{m}^2 \simeq 2 \, \text{eV}^2$
and
$\sin^2 2\vartheta \simeq 0.04$,
which is compatible with Chooz and Tritium data
(see Fig.~\ref{040}).

In Section~\ref{041}
we have discussed the tension between
the neutrino MiniBooNE and Gallium data
and
the antineutrino reactor and Tritium data.
Considering such tension as a statistical fluctuation,
we have presented the results of the combined analysis of MiniBooNE, Gallium, reactor and Tritium data:
$\Delta{m}^2 \simeq 2 \, \text{eV}^2$
and
$
0.01
\leq \sin^2 2\vartheta \leq
0.13
$
at $2\sigma$,
with best fit at
$\Delta{m}^2 \simeq 2 \, \text{eV}^2$
and
$\sin^2 2\vartheta \simeq 0.06$
(see Fig.~\ref{042}).

In Section~\ref{045},
we have presented predictions
for the effective neutrino masses in
$\beta$-decay and neutrinoless double-$\beta$-decay experiments
obtained as a consequence of the combined analysis of MiniBooNE, Gallium, reactor and Tritium data,
assuming the hierarchy of masses in Eq.~(\ref{006}).
The predicted interval for the contribution of $m_{4}$ to the effective neutrino mass
in $\beta$-decay is between about
0.06
and
0.49
eV
at $2\sigma$.
The upper part of this interval may be reached by the KATRIN experiment \cite{0910.4862}.
For neutrinoless double-$\beta$-decay we obtained a prediction for the
contribution of $m_{4}$ to the effective neutrino mass between
about
0.003
and
0.07
eV
at $2\sigma$,
which may be reached in future experiments
(see Ref.~\cite{0708.1033}).

We also considered,
in Section~\ref{058},
the possibility of reconciling the tension between
the neutrino MiniBooNE and Gallium data
and
the antineutrino reactor and Tritium data
discussed in Section~\ref{041}
with different mixings in the neutrino and antineutrino sectors.
We found a
$2.6\sigma$
indication of a mixing angle asymmetry
(99.14\% C.L.).
We pointed out the possibility of checking the mixing difference between the neutrino and antineutrino sectors
by measuring different effective electron
antineutrino and neutrino masses
in $\beta^{-}$ and $\beta^{+}$ decay experiments.

The indication in favor of short-baseline disappearance of electron neutrinos
imply the possible existence of a light sterile neutrino
which could have important consequences in
physics
\cite{hep-ph/0608147,hep-ph/0609177,hep-ph/0611178,hep-ex/0701004,0704.0388,0705.0107,0706.1462,0707.2481,0710.2985,0809.5076,0906.1997},
astrophysics
\cite{0706.0399,0709.1937,0710.5180,0712.1816,0805.4014,0806.3029,0906.2802,0906.4117,0910.5856}
and cosmology
\cite{0711.2450,0810.5133,0812.2249,0812.4552,0906.3322,1001.4440}.

As far as the effective number of neutrino species in cosmology, $N_{\text{eff}}$, is concerned,
the analysis of 7-years WMAP data has provided the following result:
$N_{\text{eff}} = 4.34^{+ 0.86}_{- 0.88}$ (68\% C.L.) \cite{Komatsu:2010fb}.
In 2011 the Planck experiment will measure $N_{\text{eff}}$ with
a factor of 4 improvement in accuracy with respect to present data \cite{Ichikawa:2008pz,1005.3808}.
In other words, the possibility of existence of a fourth light sterile neutrino
could be pursued with $5\sigma$ significance.

Finally,
we would like to encourage all experiments which can investigate the hypothesis
of short-baseline electron neutrino disappearance.

Starting from 2010, at the same $L/E$ of MiniBoone,
the magnetic off-axis near detector at
280 m of the T2K experiment \cite{Le:2009nr}
will count $\nu_e$ events with expected higher
statistics and similar $\nu_\mu$ background contamination.
A test of short-baseline oscillations may be done,
although the accuracy suffers from the scarce knowledge of the neutrino flux
and of the neutrino cross section at 1 GeV energies
\cite{0902.1992,Giganti-Longhin-private-10}.

A better measurement will be possible with the new CERN-PS neutrino beam
\cite{0909.0355,Rubbia-CERN-2010},
thanks to the presence of 2 detectors at 140 m (NEAR) and 885 m (FAR).
At $\Delta m^2 \approx 2 \, \text{eV}^2$,
the oscillation length is about 1 km for 1 GeV neutrino energies.
Therefore, one can reduce the systematic error of the Monte Carlo predictions by normalizing
the high energy part of the $\nu_{e}$ spectrum at the NEAR location.
In addition,
a better $\nu_{\mu}$ background rejection
will be possible using the liquid Argon technology.
The interesting possibility of a $\nu_e$ tagging in the CERN-PS beam
was also studied in this context \cite{Ludovici:2010ci}.

New measurements with a radioactive source could be made in the
SAGE experiment \cite{0901.2200},
with the Borexino detector
and with the future LENS detector
\cite{hep-ph/0611178}.
At $\Delta m^2 \approx 2 \, \text{eV}^2$,
the oscillation length is about 1 m for 1 MeV neutrino energies.
Therefore Borexino could measure the oscillation pattern over a distance of 4 m
(the Borexino radius) using the well known $\nu_e$-$e$ scattering process
and with a vertex resolution that at the moment is about 15 cm
\cite{Bellotti-private-10}.

At the Gran Sasso laboratories a very interesting measurement
could be realized by using the ICARUS 600 ton detector and
new low-cost and high-power proton cyclotrons under development for commercial uses
\cite{1006.0260}.
These provide electron neutrino beams with energy up to 52 MeV from muon decay-at-rest.
A low-energy $\nu_{e}$ disappearance experiment
(as well as $\boss{\nu}{\mu}$ disappearance and $\boss{\nu}{\mu}\to\boss{\nu}{e}$ measurements)
can be performed with such devices
due to the full efficiency of the ICARUS detector at 20 MeV energies.
The expected event rate is about 400 charged-current electron neutrino events per year per ton
with the ICARUS detector located at 50 m from the source \cite{1004.0310}.
The number of events is calculated assuming $10^{15}$ $\nu_e$'s per year
and a fully efficient detector.
Since at $\Delta m^2 \approx 2 \, \text{eV}^2$
the oscillation length is about 20 m for a 20 MeV neutrino energy,
it is possible to measure the full oscillation pattern along the beam direction
inside the ICARUS volume.

The disappearance of electron neutrinos
can be investigated with high accuracy in future near-detector
beta-beam \cite{0907.3145}
and
neutrino factory \cite{0907.5487,1005.3146}
experiments
in which the neutrino fluxes will be known with high precision.

Furthermore,
the MiniBooNE low-energy anomaly may be clarified
by the
ArgoNeuT,
MicroBooNE \cite{0910.3553}
and
BooNE
experiments \cite{Stancu:2009vq},
and
the magnetic off-axis near detector at
280 m of the T2K experiment has the unique opportunity to measure the charge of the events
of the low-energy anomaly \cite{Laveder-LBNL-2007}.

\section*{\label{note}Note Added}

After the completion of this work,
two important experimental results have been presented at the Neutrino 2010 conference:

\begin{enumerate}

\item
The MiniBooNE collaboration presented updated results on the search for
short-baseline $\bar\nu_{\mu}\to\bar\nu_{e}$ oscillations
which are compatible with the LSND signal
\cite{MiniBooNE-Neutrino2010,1007.1150}.
The inclusion of these data in our framework will require a
separate analysis in which the assumption of negligible $|U_{\mu4}|^2$
is relaxed \cite{Giunti-Laveder-IP-10}.

\item
The MINOS collaboration presented an indication of a possible difference between the effective mixings
of neutrinos and antineutrinos in long-baseline $\nu_{\mu}$ and $\bar\nu_{\mu}$
disappearance \cite{MINOS-Neutrino2010}.
This indication is analogous to that discussed in Section~\ref{058}.

\end{enumerate}

\begin{acknowledgments}
We would like to thank
E. Bellotti,
C. Giganti,
A. Longhin,
F. Pietropaolo
and
A. Rubbia
for interesting discussions and suggestions.
\end{acknowledgments}

%

\end{document}